\documentclass[fleqn,usenatbib]{mnras}

\usepackage{newtxtext,newtxmath}

\usepackage[T1]{fontenc}

\DeclareRobustCommand{\VAN}[3]{#2}
\let\VANthebibliography\thebibliography
\def\thebibliography{\DeclareRobustCommand{\VAN}[3]{##3}\VANthebibliography}

\usepackage{graphicx}	
\usepackage{xcolor}

\newcommand{\msol} {M$_{\odot}$}

\newcommand{\about} {$\sim$}

\definecolor{hfs_col}{HTML}{e30044}

\title[To be or not to be a black hole: BPASS as a sanity check]{To be or not to be a black hole: detailed binary population models as a sanity check}

\author[H. F. Stevance]{
H. F. Stevance,$^{1}$\thanks{E-mail: hfstevance@gmail.com}, 
S. G. Parsons$^{2}$
J. J. Eldridge$^{1}$
\\
$^{1}$Department of Physics, The University of Auckland, Private Bag 92019, Auckland, New Zealand\\
$^{2}$Department of Physics and Astronomy, The University of Sheffield, Hicks Building, S3 7RH, Sheffield, United Kingdom\\
}

\date{Accepted XXX. Received YYY; in original form ZZZ}

\pubyear{2021}

\begin{document}
\label{firstpage}
\pagerange{\pageref{firstpage}--\pageref{lastpage}}
\maketitle

\begin{abstract}
We use the self-consistent stellar populations in the Binary Population A Spectral Synthesis (BPASS) models to assess whether NGC1850-BH1 is a black hole. Using search criteria based on reported physical properties in the literature we purposefully search for suitable systems with a black hole (or compact object) companion: we  do not find any. Good  matches to the observations are found in models where the bright component is a stripped star and the companion is natively (meaning we did not impose this in our search) 1 to 2.3 magnitudes fainter than the primary in the optical bands. This alone can explain the lack of detection of the companion in the MUSE spectra without the need to invoke rapid rotation, although the conservative mass transfer exhibited by these particular models is likely to lead to rapidly rotating companions which could further smear their spectroscopic signatures.
We advise that future claims of unseen black holes in binary systems would benefit from exploring detailed binary evolution models of stellar populations as a sanity check. 
\end{abstract}




\section{Introduction}
\label{sec:intro}
Increasing the sample of stellar mass black holes (BH) with well constrained physical characteristics is essential to improving our collective understanding of stellar evolution and compact remnants. 
Gravitational wave detectors have considerably boosted our sample of BHs in binaries (e.g. \citealt{lvk2021}), but these systems are at the very end of their evolution; finding black holes in binaries prior to the formation of a second compact remnant provides further constraints on stellar evolution in intermediary stages of the life of these systems.
Recently, several studies with detailed orbital motion analysis have identified multiple systems with a BH component, such as LB1 and  HR 6819 \citep{liu2019, rivinius2020}.
But the complexity of these analyses has led to controversy surrounding the discoveries, and further work by independent groups found evidence that BH components where not responsible for the observational properties (LB1 - \citealt{irrgang2020,shenar2020, abdul-masih2020, el-badry2020}; HR6819 -\citealt{bodensteiner2020})

A new study by \cite{saracino2021} suggested the presence of a BH in the massive cluster NGC 1850 located in the Large Magellanic Cloud. 
Dubbed NGC 1850 BH1, it is reported to have a mass of 11\msol\ and be in a 5.04 day orbit around a main sequence turn-off donor star with M$_{\rm donor}$ = 4.9 \msol.
A few days later \cite{el-badry2021} showed that the period-density relationship reduced the estimated donor mass to \about 1\msol\,(with an upper limit \about2.5\msol).
They also used MESA \citep{Paxton2011, Paxton2013, Paxton2015, Paxton2018, Paxton2019} to propose an alternative binary system to explain the observations, but they emphasise that the model they present only serves as an example to show that ``more banal" alternatives are possible.

In this letter we will use the Binary Population And Spectral Synthesis (BPASS) datasets to address two questions: 1) Is the existence of NGC 1850BH1 supported by stellar evolution? 2) What type of system is most likely to explain the physical and observational properties described in \cite{saracino2021} and \cite{el-badry2021}?
BPASS includes an extensive pre-computed grid of detailed stellar evolution models which includes binary interactions -- one of the key differences between MESA and BPASS is that the former is most often used to flexibly create individual systems as needed, whereas the grid of stellar models in the later was created and tested to self-consistently reproduce a wide range of observables (see \citealt{eldridge2017, stanway2018}).
In the context of studies such as this one it is very important to emphasise that we are not using the BPASS code to create a system on the fly to recreate the observations. 
Instead we are searching through \textit{existing data products} to see if our populations can \textit{predict} observed systems. 
This gives us confidence that the models we find are consistent with stellar evolution at large, and allows us to quantify how prevalent a matched model would be in a given population. 

All the models used in this work use the BPASS fiducial initial mass function which is a \cite{kroupa2001} prescription with a maximum initial mass of 300 \msol.
Amongst the 13 metallicities available, we use Z=0.010 which is the closest match to NGC 1850. 
Finally, to facilitate searches through the vast stellar library (\about 250,000 models) we are using the freely available BPASS stellar library dataframes created with {\tt hoki} \citep{stevance2020, stevance2020d}.

In Section \ref{sec:search} we perform several searches to see if systems containing potential NGC1850-BH1 candidates are predicted, before removing the constrain of a compact companion and performing a search to identify the models that best match the observational properties of the system.
In Section \ref{sec:conclusion} we discuss the implications of our model search and conclude on the most likely nature of the "dark" companion.

\section{Model Search}
\label{sec:search}
\subsection{Looking for NGC 1850 BH1}
First let us consider the system as described by \cite{saracino2021}: a \about 5 \msol\, star with a \about 11 \msol\, black hole primary in a 5 day orbit. 
We perform a search on a wide window of parameters, selecting any secondary model with M=3--7\msol, ages between 80 and 300 Myrs, and periods ranging from 3 to 7 days. 
At this stage we do not attempt to match the observed photometry to our synthetic photometry. 
The closest match obtained are 5 \msol\, stars with black hole masses 6.3\msol, and we find no models with larger black hole masses. 
The originally described system is therefore not predicted in our simulations.

We then perform a more specific black hole (or compact remnant) search taking into account the analysis of \cite{el-badry2021}, who placed limits on the mass of the binary components.
We also include constraints on the synthetic photometry ($\pm$ 1 mag) of f336w and f814w (the two filters present in our models with reported observational values). 
The search criteria are as follows
\begin{enumerate}
    \item Must contain a compact remnant
    \item Age = [80, 120] Myrs
    \item f336w = -2.537 $\pm$ 1 mags
    \item f814w = -1.68 $\pm$ 1 mags
    \item P=5.04 $\pm$ 2 days
    \item M$_{\rm bright}$ = [0.5, 2] \msol 
    \item M$_{\rm dark}$ = [2, 6] \msol 
\end{enumerate}

The mass limits are based on the best estimates in figure 2 of \citealt{el-badry2021}.
We do not use temperature as a search criteria to allow for discussion but we come back to it at the end of this section. 
Using these constraints we find two matching models (see Table \ref{tab:bh_search}), both of which have a dark companion with a mass around the upper mass limit for a neutron star. 

\begin{table}
    \centering
    \caption{Properties of the nearest matching models with compact remnants}
    \label{tab:bh_search}
    \begin{tabular}{lllrrc}
    \hline
     MODEL  & Age  &   $P$  &    M$_{\rm bright}$ &    M$_{\rm ``dark''}$ &  $\log(T_{\rm bright}/K$) \\
     ID &  (Myrs) &  (days) &    \msol &    \msol &   \\
    \hline
       175736 &  81.10 &  4.89 &  0.90 &  2.55 &     4.81 \\
     \hline
       176986 &  80.31 &  5.42 &  1.16 &  2.26 &     4.55 \\
\hline
\end{tabular}
\end{table}

The upper mass limit of neutron stars remains debated but estimates based on the binary neutron star merger GW 170817 (see figure 3 in \citealt{abbott2018}) indicate that  Model 175736 (2.55 \msol) is very likely a black hole whereas Model 176986 (2.26\msol) may be a black hole or a neutron star depending on the equation of state considered. 
Note that the ages are lower than that of NGC 1850 by 20 Myrs, which is actually necessary for these models to be viable matches: In BPASS to save computing time the primary and secondary stars are evolved in detail successively, not simultaneously, and the age of a secondary star model \textit{is its age since the last rejuvination episode}.
A multitude of primary models could provide such systems with compact companions within 20 to 30 Myrs, and it is not within the scope of this letter to assess the most likely primary route as it does not inform the question of the nature of NGC 1850-BH1 further. 

Even without searching for the primary evolution of these systems, BPASS outputs provide an estimate of the occurrence rate of each model within a given simple stellar population. 
Models 175736 and 176986 have occurrence rates of 0.047 and 0.048 per $10^6$\msol, respectively, meaning that in NGC1850 (with M\about10$^5$\msol) there is roughly a 0.5 percent chance of the parent system occurring. 
Additionally, the phases at which the criteria are met are very short lived for each model (40 and 15,200 years, respectively), making their observation even more unlikely. 
Finally, we need to address the temperature discrepancy between the reported value in \cite{saracino2021} -- 14,500 K -- and the predicted values in our two models  -- 35,000 to 63,000K (see Table \ref{tab:bh_search}). 
Not only are our best black hole system matches extremely rare, the predicted surface temperature of the bright component actually excludes them quite confidently as viable matches to the observations.
Consequently our synthetic stellar population suggests that there is no compact remnant in NGC 1850.

But the ``dark'' companion does not have to be a compact remnant and we explore these possibilities in the next section.

\subsection{Looking for the best match}
We now search for binary models without compact remnants that match the physical properties of the system. We slightly tighten the constraint on the photometry to highlight the very best matches in our analysis. 
\begin{enumerate}
    \item Not a compact remnant companion
    \item Age = [80, 120] Myrs
    \item f336w = -2.537 $\pm$ 0.8 mags
    \item f814w = -1.68 $\pm$ 0.8 mags
    \item P=5.04 $\pm$ 2 days
    \item M$_{\rm bright}$ = [0.5, 2] \msol 
    \item M$_{\rm dark}$ = [2, 6] \msol 
    \item log(T$_{\rm bright}$) = 4 - 4.3  
\end{enumerate}

\begin{figure*}
    \centering
    \includegraphics[width=17cm]{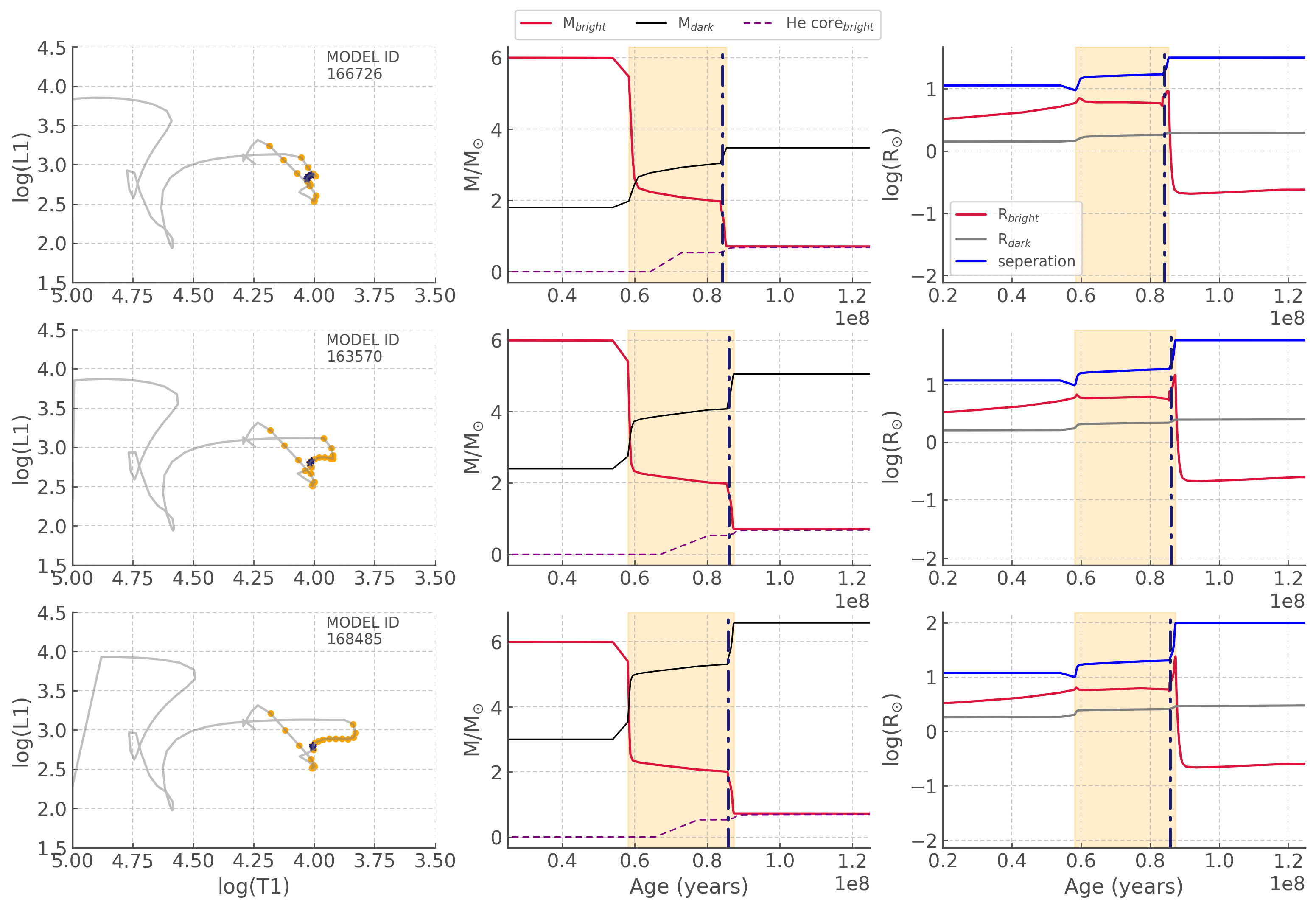}
    \caption{\textbf{Left:} Evolutionary track of the matching models in Hertzsprung-Russell Diagrams. The orange markers shows the time steps where Roche Lobe Overflow (RLOF) is occurring in the model. The dark blue markers show the steps at which the models match the search criteria. \textbf{Middle:} Evolution of the masses of both binary components. We also plot the He core mass as it shows that the RLOF occurs on the main sequence before a He core has formed (Case A). \textbf{Right:} Evolution of the separation and of the stellar radii.The shaded regions show RLOF, and the dark blue doted line shows the time at which the models match the observations.  }
    \label{fig:evolution}
\end{figure*}

We find three models (166726, 163570 and 168485) matching these criteria; their properties are summarised in Tables \ref{tab:observation_match} and \ref{tab:zams}, and their evolution is shown in Figure \ref{fig:evolution}. 
All three models are very similar: The bright star is heavily stripped with mass \about 1.27 to 1.85\msol\, and the unseen component is a 3 to 5.5 \msol\, secondary which gained mass through Roche Lobe Overflow (RLOF) starting during the Main Sequence of the primary star. 
In this scenario the system is observed at the very end of the semi-detached phase when the primary undergoes a second episode of rapid mass loss as the remainder of the the hydrogen envelope expands due to the onset of hydrogen shell burning. 
BPASS models do not record spectra of individual stars but the absolute magnitudes of both stars are predicted.
We find that in the B, V and R band (covering the wavelength range of the MUSE data obtained by \citealt{saracino2021}), the stripped primary is 1 to 2.3 magnitudes brighter than its counterpart. 
For completeness we note that in the current models the evolution of the second star is treated approximately and is not calculated in detail, and the synthetic photometry values quoted here are for a ZAMS star of the same mass (since these models have just been rejuinated through binary interaction).
In the raw BPASS data the second star is 3 to 4 magnitudes fainter than the bright component, therefore that values quoted here are conservative estimates. 
Ultimately the study of systems such as these could reveal more about the impact of mass transfer on the secondary stars of binary systems and allow us to further refine our models. 

On the whole these models correspond to rather garden variety binary stars with primary Zero Age Main Sequence masses 6 \msol\, and secondary masses \about 2 to 3 \msol\, (see Table \ref{tab:zams}).
As a result their occurrence rate is much higher than any model involving a compact remnant, with each individual system expected to occur \about1.5 times in a massive cluster such as NGC 1850 (M\about 10$^5$\msol). Collectively over 4 systems (on average) are expected to have the potential to match the observed properties in such a cluster, and the time span over which the models presented above do match our criteria ranges from 250,000 years for Model 168485 to over half a million years for Model 166726 - which means that these stars are evolving on the thermal timescale and are not unlikely to be observed in a 100 Myrs cluster such as NGC 1850.

\begin{table}
    \centering
     \caption{Properties of the stripped star models that best match observations}
    \begin{tabular}{lllrrrrr}
     \hline
     MODEL  & Age  &   $P$  &    M$_{\rm bright}$ &    M$_{\rm ``dark''}$ &  $\log$(T/K) & $\log$(g) \\
     ID &  (Myrs) &  (days) &    \msol &    \msol &   & cm.s$^{-2}$\\
    \hline
   166726 &  84.42 &  4.82917 &  1.52 &  3.24 &     4.03 & 2.84\\
   166726 &  84.61 &  5.24026 &  1.41 &  3.28 &     4.02 & 2.77 \\
   166726 &  84.76 &   5.7367 &  1.27 &  3.32 &     4.01 & 2.68 \\
   \hline
   163570 &  85.98 &  4.70969 &  1.70 &  4.33 &     4.02 & 2.89\\
   163570 &  86.26 &  5.30375 &  1.60 &  4.43 &     4.01 & 2.82\\
   \hline
   168485 &  85.79 &  4.54248 &  1.85 &  5.47 &     4.01 & 2.91\\
   168485 &  86.04 &  5.17962 &  1.74 &  5.58 &     4.01 & 2.85\\
   \hline
    \end{tabular}
    \label{tab:observation_match}
\end{table}

\begin{table}
    \centering
    \caption{Zero Age Main Sequence masses and number of such systems expected to occur per 10$^6$\msol at Z=0.010}
    \label{tab:zams}
    \begin{tabular}{llll}
\hline
        MODEL ID & M$_{\rm bright}$ &    M$_{\rm "dark"}$ & N/10$^6$\msol  \\
          \hline
        166726 &       6 &     1.8 &   2.88\\
        163570 &       6 &     2.4 &   2.78 \\
        168485 &       6 &       3 &  2.72 \\
  \hline
    \end{tabular}

\end{table}

\section{Discussion and Conclusions}
\label{sec:conclusion}
In their letter \cite{el-badry2021} suggested that NGC1850-BH1 could be a stripped star, but could not exclude the presence of a black hole as their upper mass limit on the unseen companion extended to 6\msol. 
Using the stellar populations modeled with BPASS we confirm that NGC 1850-BH1 is a stripped star, and exclude the possibility that it would be a black hole (or neutron star). 
The original interpretation that the visible star was a main sequence turn off star is understandable when using isochrones, but evolutionary tracks including binary interactions can lead stars at the end of their RLOF to circle back towards their MS turn off, which can be missed in a single-star paradigm. 

\cite{el-badry2021} suggested that the unseen companion could be hidden in the spectrum due to rapid rotation smearing the spectral lines.
In our best three matching models the stripped star is brighter than its companion in the optical bands by 1 to 2.3 magnitudes, or a factor of roughly 2.5 to >8. 
It is therefore not surprising that the companion would appear "dark" and be left undetected in the spectrum; for example, advanced disentanglement techniques were required to identify the components of the supposed LB-1 system and the final results revealed a flux contribution of 55\% and 45\% for the primary and the secondary respectively (see table 1 in \citealt{shenar2020}), whcih is a much smaller disparity than observed here. 
Overall rapid rotation likely does not need to be invoked to result in a non-detection, as suggested by \cite{el-badry2021}, but it is highly probable that the systems shown here would have rapidly rotating companions as the mass transfer episodes they underwent was rather conservative.
The systems in Models 168485, 163570 and 166726 only lost  0.18 percent, 4.7 percent and 16 percent of their pre-RLOF mass, respectively. 
Conservative mass transfer leads to the spin-up of the accretor and as a result we could expect the unseen companion to be rapidly rotating at the time the system is observed, making it harder to detect in the spectra even with a very high signal-to-noise. 
Since BPASS does not directly include rotation we are not able at this stage to quantify the rotation of the unseen companion. 

Stellar evolution models are not perfect, and population synthesis is itself limited by the fact that we can only simulate in detail a finite and discrete number of systems.
That is why the range of values applied for our search criteria must be wider than the observational errors, and although all three of our best models are the result of Case A mass transfer, it would be incorrect to conclude at this stage that the system reported in \cite{saracino2021} is \textit{necessarily} a result of Case A mass transfer. 
Additionally, the use of a discrete grid of models means that matching radial velocities directly is not a realistic endeavour, that is why we set independent mass ranges for M$_{\rm bright}$ and M$_{\rm dark}$ -- although they are rooted in the radial velocity analysis of \citealt{el-badry2021}. 
Comparison to their figure 2 indicates that Model 166726 would have very high inclination approaching 90 degrees (resulting in an eclipse, which would have been seen), whereas Models  163570 and 168485 would have inclinations around or below 60 degrees consistent with a non-eclipsing system.

Overall, there is overwhelming evidence that our stellar evolution models cannot predict NGC 1850-BH1 and that frequently occurring intermediate-mass stripped binaries are the best explanation. 
Future observational studies suggesting the presence of an unseen black hole in a binary system would benefit from using BPASS and {\tt hoki} to search for their systems in our stellar populations as a sanity check (e.g. \citealt{2020MNRAS.495.2786E}) \footnote{We have tutorials and jupyter notebooks freely available to make searching through our models as accessible as possible. We are also always happy to answer your questions on GitHub or via email.\\ \textbf{Tutorial:} \url{https://heloises.github.io/hoki/ModelSearch.html}  \\ \textbf{Further Example} -- based on \citep{stevance2021}:\\ \url{https://github.com/UoA-Stars-And-Supernovae/Binary_pathways_to_SLSNe_I_17gci}}.
If more precise matches to observations are desired, the information obtained from matching the stellar population synthesis in BPASS can then be used as a starting point to create a range of MESA models to iterate upon, thereby taking advantage of the strengths of the respective codes. 
It is also worth keeping in mind that triple systems can offer a natural explanation to quiescent BH candidates (e.g. HR6819 \citealt{romagnolo2021}) and that dynamical ineractions play a role in the formation of BH-Main Sequence star systems (e.g. \citealt{banerjee2018}).

If further observations of the NGC 1850-BH1 system confirm the existence of NGC 1850-BH1, it would pose a direct challenge to  our simulations and offer a fantastic constraint to our stellar evolution prescription, directly impacting our understanding of compact remnant pathways.

\section*{Acknowledgements}
The authors are grateful to the anonymous referee for their insightful suggestions.
We are also thankful to Chris Usher for sharing the original observational data of the system with us so quickly, as well as Sebastian Kamann and Nate Bastian for thought provoking questions in private correspondence. 
HFS and JJE acknowledge the support of the Marsden Fund Council managed through Royal Society Te Aparangi. 
SGP acknowledges the support of a Science and Technology Facilities Council (STFC) Ernest Rutherford Fellowship.
\subsection*{Third party software}
This work made extensive use matplotlib \cite{matplotlib}, numpy \citep{matplotlib, numpy} and pandas \cite{pandas1, pandas2}.

\section*{Data Availability}
The code to reproduce all the analysis and figures can be found on GitHub ({\url{https://github.com/UoA-Stars-And-Supernovae/ngc1850bh1}}) with large dataset stored on Zenodo {\url{https://zenodo.org/record/5813956\#.YdJK33VBzmE}}
The BPASSv2.2 stellar library is available as a repository of individual text (Download {\tt bpass-v2.2-newmodels.tar.gz} files\footnote{\url{https://drive.google.com/drive/folders/1BS2w9hpdaJeul6-YtZum--F4gxWIPYXl}})  or as individual tables split by metallicity\footnote{\url{https://zenodo.org/record/3905388\#.YZ603JFBzDQ}}



\bibliographystyle{mnras}
\bibliography{betterbib, extra} 

\begin{thebibliography}{}
\makeatletter
\relax
\def\mn@urlcharsother{\let\do\@makeother \do\$\do\&\do\#\do\^\do\_\do\%\do\~}
\def\mn@doi{\begingroup\mn@urlcharsother \@ifnextchar [ {\mn@doi@}
  {\mn@doi@[]}}
\def\mn@doi@[#1]#2{\def\@tempa{#1}\ifx\@tempa\@empty \href
  {http://dx.doi.org/#2} {doi:#2}\else \href {http://dx.doi.org/#2} {#1}\fi
  \endgroup}
\def\mn@eprint#1#2{\mn@eprint@#1:#2::\@nil}
\def\mn@eprint@arXiv#1{\href {http://arxiv.org/abs/#1} {{\tt arXiv:#1}}}
\def\mn@eprint@dblp#1{\href {http://dblp.uni-trier.de/rec/bibtex/#1.xml}
  {dblp:#1}}
\def\mn@eprint@#1:#2:#3:#4\@nil{\def\@tempa {#1}\def\@tempb {#2}\def\@tempc
  {#3}\ifx \@tempc \@empty \let \@tempc \@tempb \let \@tempb \@tempa \fi \ifx
  \@tempb \@empty \def\@tempb {arXiv}\fi \@ifundefined
  {mn@eprint@\@tempb}{\@tempb:\@tempc}{\expandafter \expandafter \csname
  mn@eprint@\@tempb\endcsname \expandafter{\@tempc}}}

\bibitem[\protect\citeauthoryear{Abbott et~al.,}{Abbott
  et~al.}{2018}]{abbott2018}
Abbott B.~P.,  et~al., 2018, \mn@doi [Phys. Rev. Lett.]
  {10.1103/PhysRevLett.121.161101}, 121, 161101

\bibitem[\protect\citeauthoryear{{Abdul-Masih} et~al.,}{{Abdul-Masih}
  et~al.}{2020}]{abdul-masih2020}
{Abdul-Masih} M.,  et~al., 2020, \mn@doi [\nat] {10.1038/s41586-020-2216-x},
  \href {https://ui.adsabs.harvard.edu/abs/2020Natur.580E..11A} {580, E11}

\bibitem[\protect\citeauthoryear{{Banerjee}}{{Banerjee}}{2018}]{banerjee2018}
{Banerjee} S.,  2018, \mn@doi [\mnras] {10.1093/mnras/sty2608}, \href
  {https://ui.adsabs.harvard.edu/abs/2018MNRAS.481.5123B} {481, 5123}

\bibitem[\protect\citeauthoryear{{Bodensteiner} et~al.,}{{Bodensteiner}
  et~al.}{2020}]{bodensteiner2020}
{Bodensteiner} J.,  et~al., 2020, \mn@doi [\aap] {10.1051/0004-6361/202038682},
  \href {https://ui.adsabs.harvard.edu/abs/2020A&A...641A..43B} {641, A43}

\bibitem[\protect\citeauthoryear{{El-Badry} \& Burdge}{{El-Badry} \&
  Burdge}{2021}]{el-badry2021}
{El-Badry} K.,  Burdge K.,  2021, arXiv:2111.07925 [astro-ph]

\bibitem[\protect\citeauthoryear{{El-Badry} \& {Quataert}}{{El-Badry} \&
  {Quataert}}{2020}]{el-badry2020}
{El-Badry} K.,  {Quataert} E.,  2020, \mn@doi [\mnras]
  {10.1093/mnrasl/slaa004}, \href
  {https://ui.adsabs.harvard.edu/abs/2020MNRAS.493L..22E} {493, L22}

\bibitem[\protect\citeauthoryear{Eldridge, Stanway, Xiao, McClelland, Taylor,
  Ng, Greis  \& Bray}{Eldridge et~al.}{2017}]{eldridge2017}
Eldridge J.~J.,  Stanway E.~R.,  Xiao L.,  McClelland L. A.~S.,  Taylor G.,  Ng
  M.,  Greis S. M.~L.,   Bray J.~C.,  2017, \mn@doi [Publications of the
  Astronomical Society of Australia] {10.1017/pasa.2017.51}, 34, e058

\bibitem[\protect\citeauthoryear{{Eldridge}, {Stanway}, {Breivik}, {Casey},
  {Steeghs}  \& {Stevance}}{{Eldridge} et~al.}{2020}]{2020MNRAS.495.2786E}
{Eldridge} J.~J.,  {Stanway} E.~R.,  {Breivik} K.,  {Casey} A.~R.,  {Steeghs}
  D.~T.~H.,   {Stevance} H.~F.,  2020, \mn@doi [\mnras]
  {10.1093/mnras/staa1324}, \href
  {https://ui.adsabs.harvard.edu/abs/2020MNRAS.495.2786E} {495, 2786}

\bibitem[\protect\citeauthoryear{Harris et~al.,}{Harris et~al.}{2020}]{numpy}
Harris C.~R.,  et~al., 2020, \mn@doi [Nature] {10.1038/s41586-020-2649-2}, 585,
  357

\bibitem[\protect\citeauthoryear{Hunter}{Hunter}{2007}]{matplotlib}
Hunter J.~D.,  2007, \mn@doi [Computing in Science \& Engineering]
  {10.1109/MCSE.2007.55}, 9, 90

\bibitem[\protect\citeauthoryear{{Irrgang}, {Geier}, {Kreuzer}, {Pelisoli}  \&
  {Heber}}{{Irrgang} et~al.}{2020}]{irrgang2020}
{Irrgang} A.,  {Geier} S.,  {Kreuzer} S.,  {Pelisoli} I.,   {Heber} U.,  2020,
  \mn@doi [\aap] {10.1051/0004-6361/201937343}, \href
  {https://ui.adsabs.harvard.edu/abs/2020A&A...633L...5I} {633, L5}

\bibitem[\protect\citeauthoryear{{Kroupa}}{{Kroupa}}{2001}]{kroupa2001}
{Kroupa} P.,  2001, \mn@doi [\mnras] {10.1046/j.1365-8711.2001.04022.x}, \href
  {https://ui.adsabs.harvard.edu/abs/2001MNRAS.322..231K} {322, 231}

\bibitem[\protect\citeauthoryear{{Liu} et~al.,}{{Liu} et~al.}{2019}]{liu2019}
{Liu} J.,  et~al., 2019, \mn@doi [\nat] {10.1038/s41586-019-1766-2}, \href
  {https://ui.adsabs.harvard.edu/abs/2019Natur.575..618L} {575, 618}

\bibitem[\protect\citeauthoryear{{Paxton}, {Bildsten}, {Dotter}, {Herwig},
  {Lesaffre}  \& {Timmes}}{{Paxton} et~al.}{2011}]{Paxton2011}
{Paxton} B.,  {Bildsten} L.,  {Dotter} A.,  {Herwig} F.,  {Lesaffre} P.,
  {Timmes} F.,  2011, \mn@doi [\apjs] {10.1088/0067-0049/192/1/3}, \href
  {https://ui.adsabs.harvard.edu/abs/2011ApJS..192....3P} {192, 3}

\bibitem[\protect\citeauthoryear{{Paxton} et~al.,}{{Paxton}
  et~al.}{2013}]{Paxton2013}
{Paxton} B.,  et~al., 2013, \mn@doi [\apjs] {10.1088/0067-0049/208/1/4}, \href
  {https://ui.adsabs.harvard.edu/abs/2013ApJS..208....4P} {208, 4}

\bibitem[\protect\citeauthoryear{{Paxton} et~al.,}{{Paxton}
  et~al.}{2015}]{Paxton2015}
{Paxton} B.,  et~al., 2015, \mn@doi [\apjs] {10.1088/0067-0049/220/1/15}, \href
  {https://ui.adsabs.harvard.edu/abs/2015ApJS..220...15P} {220, 15}

\bibitem[\protect\citeauthoryear{{Paxton} et~al.,}{{Paxton}
  et~al.}{2018}]{Paxton2018}
{Paxton} B.,  et~al., 2018, \mn@doi [\apjs] {10.3847/1538-4365/aaa5a8}, \href
  {https://ui.adsabs.harvard.edu/abs/2018ApJS..234...34P} {234, 34}

\bibitem[\protect\citeauthoryear{{Paxton} et~al.,}{{Paxton}
  et~al.}{2019}]{Paxton2019}
{Paxton} B.,  et~al., 2019, \mn@doi [\apjs] {10.3847/1538-4365/ab2241}, \href
  {https://ui.adsabs.harvard.edu/abs/2019ApJS..243...10P} {243, 10}

\bibitem[\protect\citeauthoryear{{Rivinius}, {Baade}, {Hadrava}, {Heida}  \&
  {Klement}}{{Rivinius} et~al.}{2020}]{rivinius2020}
{Rivinius} T.,  {Baade} D.,  {Hadrava} P.,  {Heida} M.,   {Klement} R.,  2020,
  \mn@doi [\aap] {10.1051/0004-6361/202038020}, \href
  {https://ui.adsabs.harvard.edu/abs/2020A&A...637L...3R} {637, L3}

\bibitem[\protect\citeauthoryear{{Romagnolo}, {Olejak}, {Hypki}, {Wiktorowicz}
  \& {Belczynski}}{{Romagnolo} et~al.}{2021}]{romagnolo2021}
{Romagnolo} A.,  {Olejak} A.,  {Hypki} A.,  {Wiktorowicz} G.,   {Belczynski}
  K.,  2021, arXiv e-prints, \href
  {https://ui.adsabs.harvard.edu/abs/2021arXiv210708930R} {p. arXiv:2107.08930}

\bibitem[\protect\citeauthoryear{Saracino et~al.,}{Saracino
  et~al.}{2021}]{saracino2021}
Saracino S.,  et~al., 2021, arXiv:2111.06506 [astro-ph]

\bibitem[\protect\citeauthoryear{{Shenar} et~al.,}{{Shenar}
  et~al.}{2020}]{shenar2020}
{Shenar} T.,  et~al., 2020, \mn@doi [\aap] {10.1051/0004-6361/202038275}, \href
  {https://ui.adsabs.harvard.edu/abs/2020A&A...639L...6S} {639, L6}

\bibitem[\protect\citeauthoryear{Stanway \& Eldridge}{Stanway \&
  Eldridge}{2018}]{stanway2018}
Stanway E.~R.,  Eldridge J.~J.,  2018, \mn@doi [Monthly Notices of the Royal
  Astronomical Society] {10.1093/mnras/sty1353}, 479, 75

\bibitem[\protect\citeauthoryear{{Stevance} \& {Eldridge}}{{Stevance} \&
  {Eldridge}}{2021}]{stevance2021}
{Stevance} H.~F.,  {Eldridge} J.~J.,  2021, \mn@doi [\mnras]
  {10.1093/mnrasl/slab039}, \href
  {https://ui.adsabs.harvard.edu/abs/2021MNRAS.504L..51S} {504, L51}

\bibitem[\protect\citeauthoryear{Stevance, Eldridge  \& Stanway}{Stevance
  et~al.}{2020a}]{stevance2020}
Stevance H.,  Eldridge J.,   Stanway E.,  2020a, \mn@doi [The Journal of Open
  Source Software] {10.21105/joss.01987}, 5, 1987

\bibitem[\protect\citeauthoryear{{Stevance}, {Eldridge}  \&
  {Stanway}}{{Stevance} et~al.}{2020b}]{stevance2020d}
{Stevance} H.,  {Eldridge} J.,   {Stanway} E.,  2020b, \mn@doi [The Journal of
  Open Source Software] {10.21105/joss.01987}, \href
  {https://ui.adsabs.harvard.edu/abs/2020JOSS....5.1987S} {5, 1987}

\bibitem[\protect\citeauthoryear{{The LIGO Scientific Collaboration}
  et~al.,}{{The LIGO Scientific Collaboration} et~al.}{2021}]{lvk2021}
{The LIGO Scientific Collaboration} et~al., 2021, arXiv e-prints, \href
  {https://ui.adsabs.harvard.edu/abs/2021arXiv211103634T} {p. arXiv:2111.03634}

\bibitem[\protect\citeauthoryear{{W}es {M}c{K}inney}{{W}es
  {M}c{K}inney}{2010}]{pandas2}
{W}es {M}c{K}inney 2010, in {S}t\'efan van~der {W}alt {J}arrod {M}illman eds,
  {P}roceedings of the 9th {P}ython in {S}cience {C}onference. pp 56 -- 61,
  \mn@doi{10.25080/Majora-92bf1922-00a}

\bibitem[\protect\citeauthoryear{pandas~development team}{pandas~development
  team}{2020}]{pandas1}
pandas~development team T.,  2020, pandas-dev/pandas: Pandas,
  \mn@doi{10.5281/zenodo.3509134}, \url
  {https://doi.org/10.5281/zenodo.3509134}

\makeatother
\end{thebibliography}





\bsp	
\label{lastpage}
\end{document}